\documentclass[letter]{aa}  
\usepackage{graphicx}
\usepackage{txfonts}
\newcommand{\kms}{km\,s$^{-1}$}

\begin{document}

   \title{Organic molecules in the protoplanetary disk of DG Tau \\ revealed by ALMA}

   \author{L. Podio
          \inst{1}
          \and
F. Bacciotti
          \inst{1}
          \and
D. Fedele
         \inst{1}
          \and
C. Favre
          \inst{1}
          \and
C. Codella
          \inst{1,2}
          \and
K. L. J. Rygl
          \inst{3}
          \and
I. Kamp
          \inst{4}
          \and
G. Guidi
          \inst{5}
          \and
E. Bianchi
          \inst{2}
          \and
C. Ceccarelli
          \inst{2}
          \and
D. Coffey
          \inst{6,7}
          \and
A. Garufi
          \inst{1}
          \and
L. Testi
          \inst{8,1}
          }

   \institute{INAF - Osservatorio Astrofisico di Arcetri, Largo E. Fermi 5, 50125 Firenze, Italy\\
              \email{lpodio@arcetri.astro.it}
         \and Univ. Grenoble Alpes, IPAG, F-38000 Grenoble, France
         \and INAF–Istituto di Radioastronomia \& Italian ALMA Regional Centre, via P. Gobetti 101, 40129 Bologna, Italy
         \and Kapteyn Astronomical Institute, University of Groningen, Landleven 12, 9747 AD Groningen, The Netherlands
         \and ETH Zurich, Institute for Particle Physics and Astrophysics, Wolfgang-Pauli-Str. 27, 8093 Zurich, Switzerland
         \and School of Physics, University College Dublin, Belfield, Dublin 4, Ireland
         \and School of Cosmic Physics, The Dublin Institute for Advanced Studies, Dublin 2, Ireland
         \and European Southern Observatory (ESO), Karl-Schwarzschild-Str. 2, 85748, Garching, Germany
             }

   \date{Received ...; accepted ...}

 
  \abstract
   {Planets form in protoplanetary disks and inherit their chemical compositions.}
   {It is thus crucial to map the distribution and investigate the formation of simple organics, such as formaldehyde and methanol, in protoplanetary disks.}
   {We analyze ALMA observations of the nearby disk-jet system around the T Tauri star DG Tau in the o-H$_2$CO $3_{1,2}-2_{1,1}$ and CH$_3$OH $3_{-2,2}-4_{-1,4}$ E, $5_{0,5}-4_{0,4}$ A transitions at an unprecedented resolution of $\sim 0\farcs15$, i.e., $\sim 18$ au at a distance of 121 pc.}
   {The H$_2$CO emission originates from a rotating ring extending from $\sim 40$ au with a peak at $\sim 62$ au, i.e., at the edge of the 1.3~mm dust continuum. CH$_3$OH emission is not detected down to an r.m.s. of 3 mJy/beam in the 0.162 \kms\, channel. Assuming an ortho-to-para ratio of $1.8-2.8$ the ring- and disk-height-averaged H$_2$CO column density is $\sim 0.3-4 \times 10^{14}$ cm$^{-2}$, while that of CH$_3$OH is $<0.04-0.7 \times 10^{14}$ cm$^{-2}$. In the inner $40$ au no o-H$_2$CO emission is detected with an upper limit on its beam-averaged column density of  $\sim 0.5-6 \times 10^{13}$ cm$^{-2}$. }
{The H$_2$CO ring in the disk of DG Tau is located beyond the CO iceline (R$_{\rm CO} \sim 30$~au). This suggests that the H$_2$CO abundance is enhanced in the outer disk due to formation on grain surfaces by the hydrogenation of CO ice. The emission peak at the edge of the mm dust continuum may be due to enhanced desorption of H$_2$CO in the gas phase caused by increased UV penetration and/or temperature inversion.
The CH$_3$OH/H$_2$CO abundance ratio is $< 1$, in agreement with disk chemistry models. 
The inner edge of the H$_2$CO ring coincides with the radius where the polarization of the dust continuum changes orientation, hinting at a tight link between the H$_2$CO chemistry and the dust properties in the outer disk and at the possible presence of substructures in the dust distribution.}


   \maketitle
%

\begin{table*}
  \caption[]{\label{tab:lines} Properties of the observed lines.}
  \begin{tabular}[h]{cccccccc}
    \hline
    \hline
    Line & $\nu_{0}$ & E$_{\rm up}$ & S$_{ij} \mu^2$ & HPBW (PA) & r.m.s.        &  F$_{\rm int}$        & N$_{\rm X}$ \\
            & (MHz)      & (K)              &    (D$^2$)        &                 & (mJy/beam) & (mJy \kms)  &  ($10^{14}$ cm$^{-2}$)\\
    \hline 
o-H$_2$CO $3_{1,2}-2_{1,1}$      & 225697.775 & 33  & 43.5& $0\farcs17 \times 0\farcs13$ ($-20\degr$) & 1.7 & 292      & $0.2-2.7$\\
CH$_3$OH $3_{-2,2}-4_{-1,4}$ E & 230027.060 & 40 & 0.7 & $0\farcs16 \times 0\farcs13$ ($-21\degr$)& 3    & $<1.6$ & $<0.4-10$\\
CH$_3$OH $5_{0,5}-4_{0,4}$ A & 241791.431 & 35 & 4    & $0\farcs16 \times 0\farcs12$ ($-21\degr$)& 3     & $<0.7$  & $<0.04-0.7$\\
    \hline     
  \end{tabular}
\end{table*}


\section{Introduction}

A key open question in astrochemistry is how chemical complexity increases during the formation process of Sun-like stars from prestellar cores to protoplanetary disks and ultimately to planets \citep{caselli12a}. 
Is the chemical composition of planets inherited from the prestellar and protostellar stages?
Or does it reflect chemical processes occurring in the disk?
Are organics efficiently formed in disks and by what mechanism(s)?

Organic and prebiotic molecules  form either through gas-phase reactions \citep[e.g.,][]{millar91,balucani15} or on the icy surface of dust grains \citep[e.g.,][]{tielens82,garrod08}. Understanding the efficiency and occurrence of these mechanisms requires a comparison of observations and predictions from astrochemical models.
Detailed models of the disk chemistry including gas-phase reactions, molecule freeze-out, dust surface chemistry, and both thermal and nonthermal desorption mechanisms have been developed to predict the ice and gas abundances of complex organic molecules \citep[e.g.,][]{aikawa99,willacy09,walsh14,loomis15}. 
However, only a few simple organics have been detected in disks so far \citep{qi13a,oberg15,walsh16,favre18}. Among these organics, formaldehyde (H$_2$CO) and methanol (CH$_3$OH) are essential to the investigation of the formation of  organics. While H$_2$CO can form both in the gas phase and on grains, CH$_3$OH forms exclusively on grains \citep[e.g.,][]{watanabe02}. 
Therefore, observations of its abundance and distribution in disks is crucial to constraining the mechanism(s) that form these simple organics, which are the  building blocks  in the formation of complex organic and prebiotic molecules.

Observationally, H$_2$CO has been detected in a number of protoplanetary disks through single-dish and low angular resolution surveys \citep[e.g.,][]{oberg10,oberg11,guilloteau13,qi13a}, but only three of them were imaged at an angular resolution of $\sim 0.5''$, namely DM Tau \citep{loomis15}, TW Hya \citep{oberg17}, and HD 163296 \citep{carney17}. Therefore, the distribution and formation mechanisms (whether in the gas phase or on grains) of H$_2$CO in disks is still unclear.
Because of its low volatility and large partition function, CH$_3$OH is even more difficult to observe. To date, it has been detected only in the disks of TW Hya \citep{walsh16} and of the young outbursting star V883 Ori \citep{vanthoff18}.

The nearby T Tauri star DG Tau ($d=121\pm2$ pc, \citealt{gaia16,gaia18}) is an ideal target to investigate the origin of simple organics in protoplanetary disks.
DG Tau is surrounded by a compact and massive dusty disk imaged with CARMA \citep{isella10} and ALMA in polarimetric mode \citep{bacciotti18}, and is associated with bright molecular emission detected with the IRAM 30m telescope and {\it Herschel} \citep{guilloteau13,podio12,fedele13}. 
However, the origin of the detected molecular emission is unclear because DG Tau is also associated with a residual envelope and a jet \citep{eisloffel98}.
\citet{guilloteau13} suggest that the single-peaked profile of SO and H$_2$CO is due to envelope emission, while the fundamental H$_2$O lines have double-peaked profiles and fluxes in agreement with disk model predictions \citep{podio13}.
Interferometric maps of CO and its isotopologues show that the envelope dominate the molecular emission on large scales \citep{schuster93,kitamura96a}, while disk emission is detected on scales $<2"$ \citep{testi02,gudel18}.
Given the complexity of the circumstellar environment, spatially resolved maps are crucial to reveal the origin of the molecular emission.

In this letter we present ALMA Cycle 4 observations of H$_2$CO and CH$_3$OH in the disk of DG Tau at an unprecedented resolution of $\sim 0\farcs15$, i.e., $\sim 18$ au, and we discuss the possible formation mechanism of these simple organics. 

\section{Observations}

ALMA observations of DG\,Tau were performed during Cycle~4 in August 2017 with baselines ranging from 17\,m to 3.7\,km. The bandpass was calibrated with the quasar J0510+1800, and phase calibration was performed every $\sim$8 minutes using quasar J0438+3004.
The correlator setup consists of 12 high-resolution (0.122\,MHz) spectral windows (SPWs) covering several molecular transitions, among which o-H$_2$CO $3_{1,2}-2_{1,1}$, and CH$_3$OH $3_{-2,2}-4_{-1,4}$ E  and $5_{0,5}-4_{0,4}$ A (frequency, $\nu_{0}$; upper level energies, E$_{\rm up}$; and line strengths, S$_{\rm ij} \mu^2$ are listed in Table~\ref{tab:lines}). 
Data reduction was carried out following standard procedures using the ALMA pipeline in CASA 4.7.2. Self-calibration was performed on the source continuum emission, by combining a selection of line-free channels, and applying the phase solutions to the continuum-subtracted SPWs.
Continuum images and spectral cubes were produced with tclean applying a manually selected mask and Briggs parameter of 0.5. 
The flux calibration was performed using quasars J0238+1636 and J0510+1800, obtaining an accuracy of $\sim 10\%$.
The clean beam FWHMs and r.m.s (per 0.162 \kms\, channel) of the resulting line cubes are listed in Table~\ref{tab:lines}.
The continuum image has an r.m.s. of 0.3 mJy/beam.

   \begin{figure*}
   \centering
   \includegraphics[width=8.5cm]{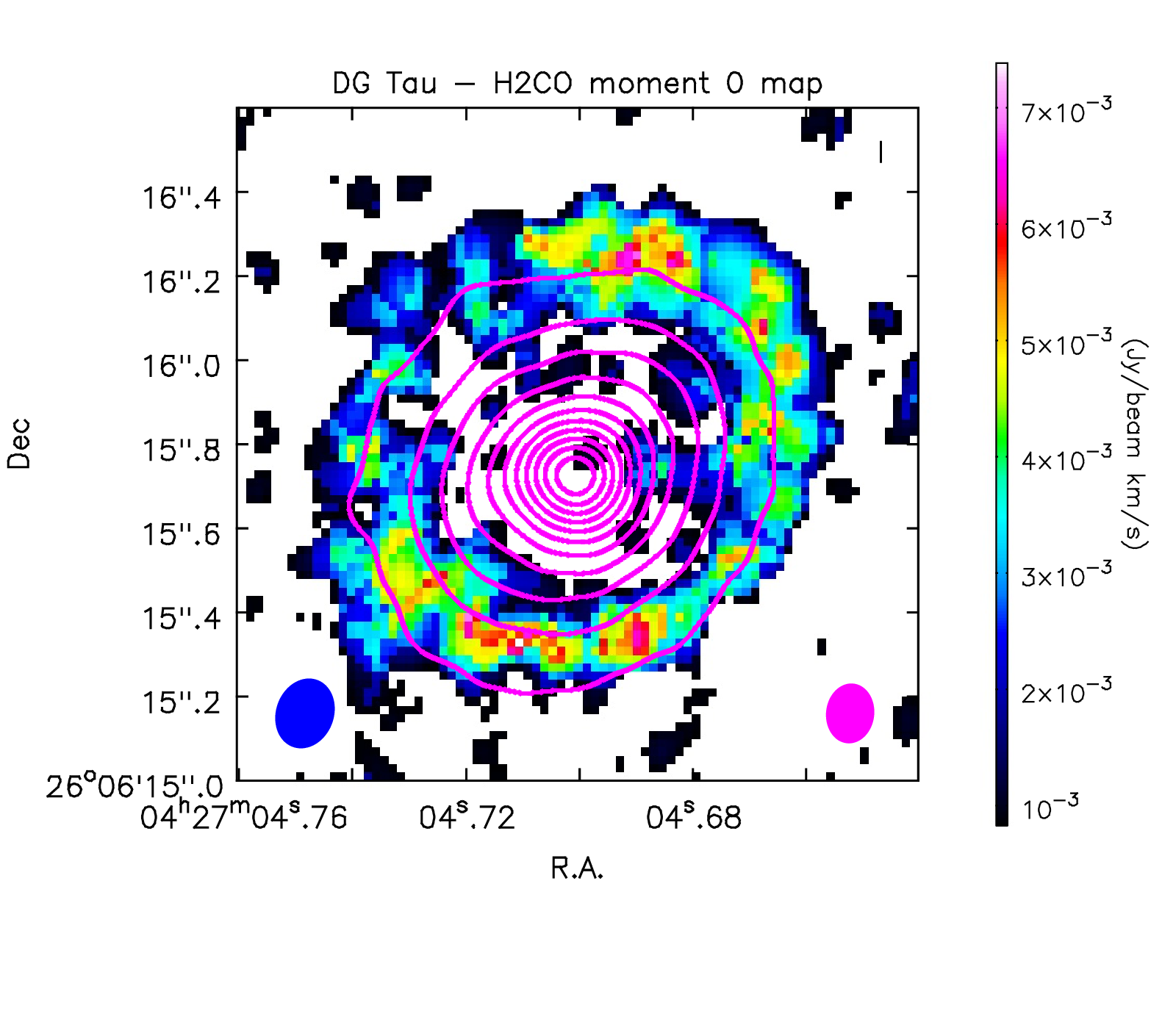}
\hspace{1cm}
   \includegraphics[width=8.5cm]{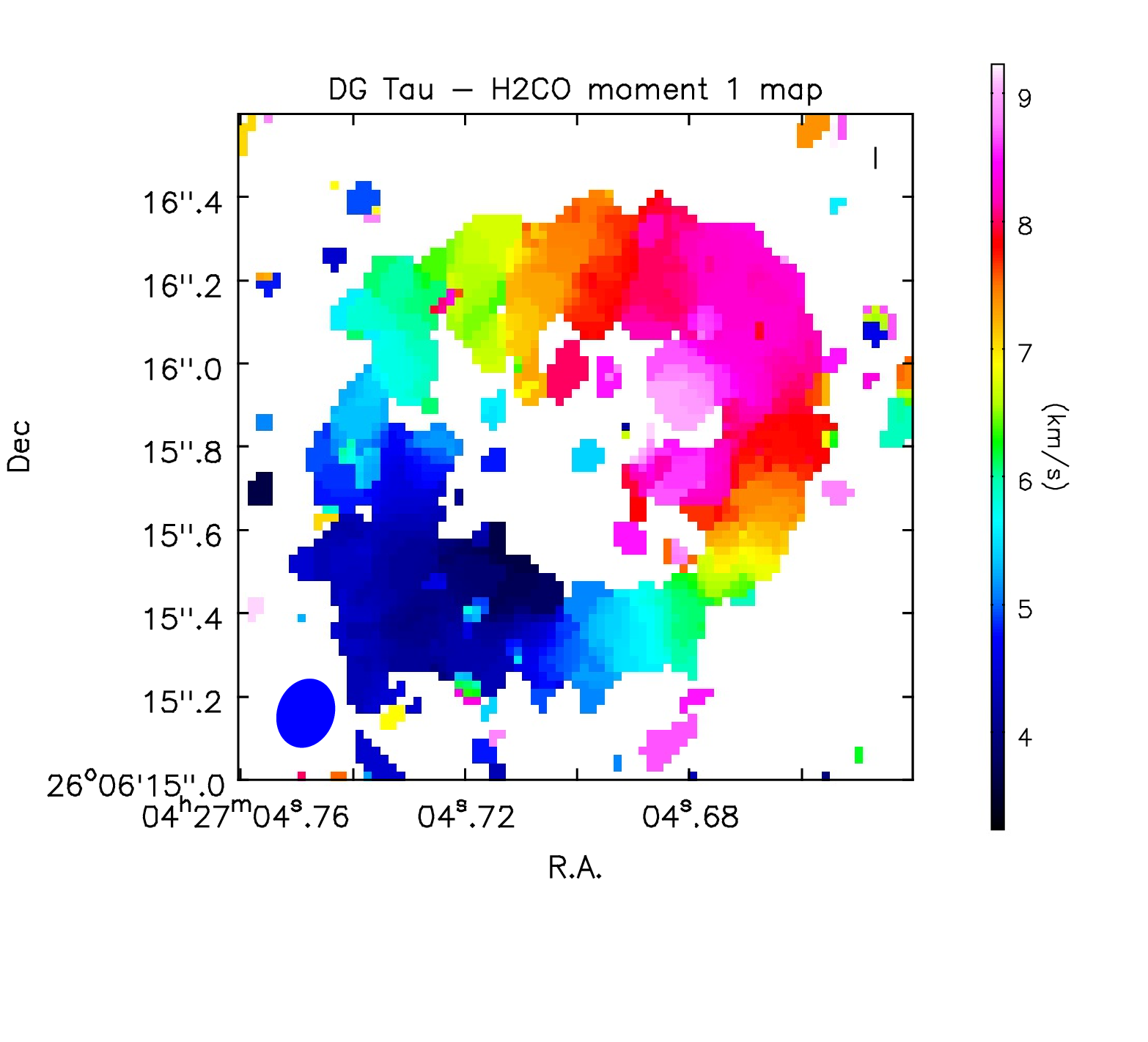}
   \vspace{-0.8cm}
   \caption{Moment maps of H$_2$CO  towards DG Tau. {\em Left panel:} H$_2$CO moment 0 map. The color scale indicates the line intensity integrated over the H$_2$CO velocity profile (V$_{\rm LSR} = (+3.24, +9.24)$ \kms) in Jy/beam \kms. The magenta contours indicate the 1.3~mm dust continuum emission (from 3$\sigma$ with steps of 10$\sigma$). The blue and magenta ellipse in the bottom left and right corner indicate the synthesized beam for H$_2$CO and the continuum, respectively. {\em Right panel:} H$_2$CO moment 1 map. The color scale indicates the velocity V$_{\rm LSR}$ in \kms.} 
              \label{fig:moments}
    \end{figure*}

\section{Ring of formaldehyde in the disk of DG Tau}

The map of the continuum emission at 225.7 GHz (1.3~mm) (Fig.~\ref{fig:moments}) shows a smooth distribution, with integrated flux of $297$ mJy and peak intensity of $33$ mJy/beam at R.A.(J2000)$=04^h 27^m 04^s.70$, Dec(J2000)$=+26\degr 06\arcmin 15\farcs72$.
By applying a 2D Gaussian fit, we determine the disk major and minor axes deconvolved from the beam ($0\farcs43\pm0\farcs02$ and $0\farcs38\pm0\farcs02$), hence the disk inclination $i= 41\degr \pm 2\degr$ and position angle PA$ = 128\degr \pm 16\degr$. These values are in good agreement with the estimates obtained at 0.87~mm and 1.3~mm \citep{bacciotti18,isella10} and with the jet inclination and PA (the jet is ejected perpendicular to the disk at $i_{\rm jet}= 38\degr \pm 2\degr$, PA $_{\rm jet} =226\degr \pm 10\degr$, \citealt{eisloffel98}).

The moment 0 and 1 maps of the H$_2$CO $3_{1,2} - 2_{1,1}$ emission obtained with CASA by integrating over velocities from $3.24$ to $9.24$ \kms, and masking all pixels where no emission is detected above the $3\sigma$ level are shown in Fig. \ref{fig:moments}, while the moment 8 map obtained without applying a threshold is shown in the Appendix (see Fig.~\ref{fig:mom8}).
Then radial intensity profiles of the H$_2$CO $3_{1,2} - 2_{1,1}$ line and of the 1.3~mm continuum are obtained by azimuthally averaging the intensity maps after deprojecting for $i= 41\degr$ and PA$ = 128\degr$ (see Fig. \ref{fig:profiles}).
The H$_2$CO $3_{1,2} - 2_{1,1}$ moment maps show an H$_2$CO ring rotating around the central star which extends from $\sim 40$ au to beyond the edge of the 1.3~mm dust emission (R$_{\rm 1.3~mm} = 66$ au at 3$\sigma$) (see also the channel maps in Fig.~\ref{fig:channel_maps}). The intensity profile in Fig.~\ref{fig:profiles} shows that the peak of the H$_2$CO emission is at $\sim 62$ au ($\sim 6.5$ mJy/beam \kms). 
The 1D spectrum integrated over a $0\farcs33 - 0\farcs75$ ring area is centered at the systemic velocity $V_{\rm sys} = + 6.24$ \kms\, and shows a double-peaked profile (see Fig. \ref{fig:1dspec}).  Assuming Keplerian rotation around a $0.7$ M$_{\odot}$ star as in \citet{podio13}\footnote{The stellar mass obtained from the pre-main sequence tracks by \citet{siess00} is not affected by the change in distance ($140$ pc in previous studies) as the tracks are vertical in this region of the HR diagram.} and $i=41\degr$, the H$_2$CO peaks at V$_{\rm LSR} - V_{\rm sys}= \pm 1.95$ \kms\, indicate an emitting radius of $\sim 70$ au, in agreement with the position of the H$_2$CO ring.

The H$_2$CO line peak and integrated flux collected with the 11$\arcsec$ beam of the IRAM 30m telescope \citep{guilloteau13} is about a factor of four larger than that obtained integrating the ALMA cube over an $11\arcsec$ circular area. This indicates that ALMA filters out the extended H$_2$CO emission from the outflow or the envelope, thus isolating the compact H$_2$CO emission from the disk.

\section{Abundance of organic molecules}
\label{sect:abu}

To estimate the H$_2$CO column density, the emission is integrated over a circular ring from $0\farcs33$ to $0\farcs75$ ($40-90$ au) and over the line velocity profile (between $+3.24$ and $+9.24$ \kms).
The maps of the undetected CH$_3$OH lines are integrated over the same spatial region and velocity range as for the H$_2$CO line (i.e., the same pixels in R.A., Dec, and V$_{\rm LSR}$) to recover an upper limit on the methanol column density  under the assumption that CH$_3$OH originates from the same disk region as H$_2$CO. 
The ring-averaged o-H$_2$CO and CH$_3$OH column densities, N$_{\rm o-H_2CO}$ and N$_{\rm CH_3OH}$, are derived from the integrated line intensities by assuming local thermodynamic equilibrium (LTE) and optically thin emission \citep[see, e.g., Eqs. (1) and (2) in][]{bianchi17a}, and adopting the molecular parameters and partition function from the Cologne Database of Molecular Spectroscopy (CDMS, \citealt{muller01}). The assumption of LTE is clearly justified as the gas density near the midplane and in the intermediate disk layers where H$_2$CO is thought to originate (see, e.g,  \citealt{walsh14,oberg17}, and the discussion in Sect.~\ref{sect:formation}) is high (from $\sim 10^{8}$ to $\sim 10^{12}$ cm$^{-3}$ according to the disk model by \citealt{podio13}), i.e., well above the critical density of the H$_2$CO $3_{1,2} - 2_{1,1}$ transition (n$_{\rm cr}\sim 7-4.6 \times 10^5$ cm$^{-3}$ at $20-100$ K, \citealt{shirley15}). The gas temperature in these intermediate disk layers is between 30 and 300 K according to the same disk model. Therefore, assuming an excitation temperature T$_{\rm ex} = 30-300$ K we infer an average o-H$_2$CO column density over the ring and the disk height of $\sim 0.2-2.7 \times 10^{14}$ cm$^{-2}$.  Following the same procedure as for H$_2$CO, the most stringent constraint on the CH$_3$OH column density is derived from the CH$_3$OH 241.791 GHz line (which has the greatest line strength S$_{\rm ij} \mu^2 \sim 4$ D$^2$): N$_{\rm CH_3OH} < 0.04-0.7 \times 10^{14}$ cm$^{-2}$. The inferred column densities are summarized in Table \ref{tab:lines}.
The total H$_2$CO column density is then obtained by assuming an ortho-to-para ratio of $1.8-2.8$ \citep{guzman18a}: N$_{\rm H_2CO}\sim 0.3-4 \times 10^{14}$ cm$^{-2}$.
Using the H$_2$ column density in the outer disk ($> 40$ au) from the disk model by \citet{podio13} (N$_{\rm H} \sim 0.3-1 \times 10^{25}$ cm$^{-2}$), we estimate the  ring- and disk-height-averaged abundance of organic molecules: X$_{\rm H_2CO} \sim 6 \times 10^{-12} - 3 \times 10^{-10}$,  X$_{\rm CH_3OH} < 0.8 \times 10^{-12} - 5 \times 10^{-11}$.
Finally, in the inner $40$ au no o-H$_2$CO is detected above 3$\sigma$.
Taking into account the wider velocity range over which the emission may be spread, this provides an upper limit on the o-H$_2$CO beam-averaged column density of $\sim 0.5-6 \times 10^{13}$ cm$^{-2}$.

The inferred H$_2$CO column density and abundance, and the upper limits retrieved for CH$_3$OH are in agreement with the predictions of chemistry models of protoplanetary disks around T Tauri stars (see, e.g., \citealt{walsh14}). These models predict gas-phase fractional abundances in the disk outer regions ($r>50$ au) of between $10^{-13}$ and a few  $10^{-10}$ for H$_2$CO and CH$_3$OH.
The H$_2$CO abundance is in agreement within a factor of a few with that estimated in the disk of TW Hya and HD 163296 assuming T$_{\rm ex} = 25$ K \citep{carney19},
while it is  up to two orders of magnitude lower than that found in disks with large inner holes, for example Oph IRS48 and AB Aur where the outer disk reservoir of icy organics is directly exposed to stellar illumination \citep{vandermarel14,pachecovazquez16}.
Also, the  nondetection of the targeted methanol transitions is consistent with a CH$_3$OH/H$_2$CO ratio $< 1$, in agreement with the prediction of chemical models  \citep[e.g.,][]{walsh14} and with the upper limit found in the disk of HD 163296 (CH$_3$OH/H$_2$CO $< 0.24$), while it is lower than that estimated in the disk of TW Hya (CH$_3$OH/H$_2$CO $= 1.27$) \citep{carney19}.

\section{H$_2$CO formation}
\label{sect:formation}

Disk chemistry models by \citet{willacy09}, \citet{walsh14}, and \citet{loomis15} show that H$_2$CO can form either in the gas phase, primarily through the reaction CH$_3$ + O; or on dust grains via hydrogenation of CO locked in their icy mantles (CO + H --> HCO + H -> H$_2$CO).
The gas-phase CH$_3$ + O formation route is efficient in the warm inner region and upper layers of the disk where atomic oxygen is produced by photodissociation of gas-phase CO.
On the contrary, formation on grains occurs in the cold outer disk beyond the CO iceline where CO is condensed onto grains and available for H$_2$CO formation.

\begin{figure}
   \centering
\vspace{-1cm}
\includegraphics[width=8.5cm]{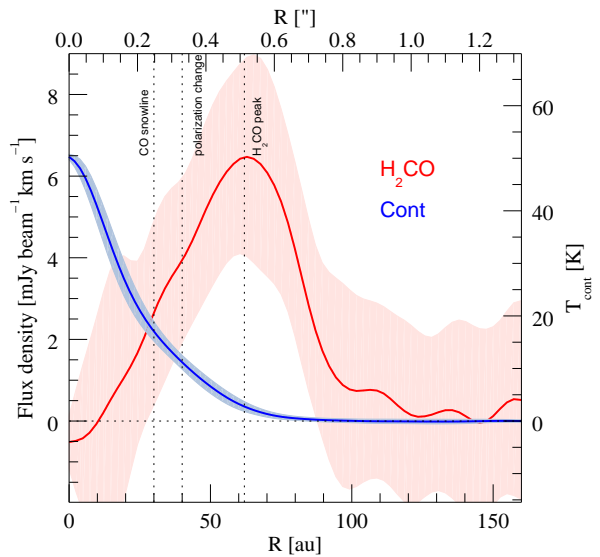}
\vspace{-0.3cm}
   \caption{Azimuthally averaged radial intensity profile of H$_2$CO $3_{1,2}-2_{1,1}$ (in mJy beam$^{-1}$ \kms, red line) and of the 1.3~mm continuum (in K, blue line).
For the continuum the conversion factor from brightness temperature to flux density is 0.65 mJy beam$^{-1}$/K.
The shaded areas indicate the dispersion of the intensity values around the mean along each annulus in the radial direction. The vertical black dotted lines indicate the position of the CO iceline, the change in the polarization orientation, and the H$_2$CO emission peak.}
              \label{fig:profiles}
    \end{figure}

Resolved observations of H$_2$CO emission allow us to constrain the formation mechanism of H$_2$CO in disks.
The H$_2$CO radial intensity profile in the disk of DG Tau (Fig.~\ref{fig:profiles}) shows characteristics in common with the three other disks observed at high angular resolution with ALMA (DM Tau, \citealt{loomis15}; TW Hya, \citealt{oberg17}; and HD 163296, \citealt{carney17}):
(i) the depression (or lack) of emission in the inner disk;
(ii) the emission peak located outside the CO iceline;
(iii) emission beyond the mm dust continuum (with a peak at the edge of the continuum in the case of DG Tau, TW Hya, and HD 163296).

In the case of the HD 163296 disk, it was argued that the central depression of the H$_2$CO emission is caused by absorption by optically thick dust continuum \citep{carney17}, while \citet{oberg17}
explain it as a real drop in abundance in TW Hya.
For DG Tau, we can exclude problems due to continuum subtraction as the line-free channels of the continuum subtracted cube do not show negative flux values in the central regions.
To evaluate if absorption by the dust continuum affects the observed H$_2$CO intensity, we estimate the optical depth of the disk at 1.3~mm from the dust opacity and surface density given by \citet{isella10}. According to their modeling, the disk is optically thick in the innermost region ($\tau_{\rm 1.3~mm} > 1$ at $< 3$~au or $<9$~au, depending on whether the self-similarity or power-law parametrization for the surface density is adopted). Then the optical depth rapidly drops to $\sim 0.5$ at 20 au and $<0.3$ for $r>30$ au\footnote{At the adopted distance of 121 pc the radius where the disk becomes optically thin is  $\sim 15\%$ smaller with respect to the modeling by \citet{isella10} who adopted d=140 pc.}. 
On the other hand, observations of $^{13}$CO $2-1$ and C$^{18}$O $2-1$ show a hole in the inner $25$ au, which could be due to the thickness of the continuum up to this radius \citep{gudel18}. 
To further check the continuum optical depth we compared its brightness temperature (T$_{\rm cont}$, see Fig.~\ref{fig:profiles}) to the expected dust temperature T$_{\rm dust}$
by \citet{podio13} (their Fig. 2, where T$_{\rm gas} = {\rm T}_{\rm dust}$ in the midplane).  We found T$_{\rm cont}/{\rm T}_{\rm dust} <0.6$, hence $\tau < 1$, for $r > 25$ au, which indicates that the continuum should be optically thin beyond this radius, in agreement with the observed distribution of $^{13}$CO $2-1$ and C$^{18}$O $2-1$.
Finally, we checked the  H$_2$CO $3_{1,2} - 2_{1,1}$ line opacity by using the non-LTE radiative transfer code RADEX \citep{vandertak07}.  Assuming column densities up to N$_{\rm H_2CO} \sim 10^{15}$ cm$^{-2}$ in the inner disk region ($10-40$ au) \citep{walsh14}, a line width of $\sim 6$ \kms, and gas densities and temperatures from our ProDiMo DG Tau model (n$_{\rm H_2} \sim 10^{10} - 10^{12}$ cm$^{-3}$ and T$_{\rm gas} > 30$ K at $10-40$ au and intermediate disk height), the H$_2$CO line optical depth reach a maximum value of $\sim 1$.
Therefore, we conclude that
the H$_2$CO ring is related to an increase in the H$_2$CO abundance beyond $\sim 40$ au.

As shown by \citet{loomis15}, disk models that include only H$_2$CO gas-phase formation produce centrally peaked column density and emission profiles, and strongly underestimate the observed H$_2$CO emission in the outer disk. H$_2$CO formation on grains followed by desorption in the gas phase  has to be taken into account to increase the H$_2$CO column densities and line emission in the outer disk (by up to two orders of magnitude in the case of DM Tau).
The intensity profile of H$_2$CO  in the disk of DG Tau shows a steep increase in the outer disk with an emission peak at $62$ au.
Interestingly, the thermo-chemical ProDiMo disk model presented in \citet{podio13}, which matches the long wavelength part of the SED of DG Tau and the {\it Herschel}/HIFI water lines, shows that the CO iceline, R$_{\rm CO}$, is located at $\sim30$ au\footnote{R$_{\rm CO}$ is derived from a steady-state solution of a chemical network including gas-phase chemistry as well as adsorption and (non)thermal desorption processes \citep[e.g.,][]{kamp10} for a stellar luminosity $L_{*}\sim 1$ L$_{\odot}$ at 140 pc. At the adopted distance of 121 pc, $L_{*}$ decreases by $\sim 25\%$, hence the estimated R$_{\rm CO}$ is an upper limit.}.
The fact that the peak is located outside the CO iceline suggests that the outer H$_2$CO reservoir is mainly produced by CO hydrogenation on grains.
This is in agreement with disk chemistry models and with observations by \citet{loomis15}, \citet{oberg17}, and \citet{carney17}. 
Contrary to the disks of DM Tau, TW Hya, and HD 163296, in the case of DG Tau no H$_2$CO emission is detected in the inner disk, where gas-phase formation dominates. 
As discussed above, this could be due to absorption by the optically thick dust continuum (up to $\sim25$ au) and to a lower H$_2$CO abundance (by a factor of  $\le 10$)  and column density in the inner 40 au (see Sect.~\ref{sect:abu}). Also taking into account  the wider velocity range over which the emission may be spread, this would produce no detectable emission in the present observations (r.m.s. $\sim 1.7$ mJy/beam). 
Finally, the fact that the H$_2$CO emission peaks at the edge of the mm dust continuum and extends beyond it could be due to a steeper decrease in the density of the gas and of the small grains at this location. As small grains are responsible for the opacity at UV wavelengths, the UV radiation penetrates deeper in the disk and more efficiently photodesorbs H$_2$CO from grains \citep{oberg15b,oberg17}. Moreover, \citet{cleeves16} suggests that there could be a temperature inversion at the edge of the large dust population, which would enhance thermal desorption.  Alternatively, H$_2$CO could be efficiently formed in the gas phase beyond the mm-dust edge due to more efficient photodissociation of CO favoring the CH$_3$ + O formation route \citep{carney17}.
An enhancement of molecular emission at the edge of the millimeter dust continuum has been observed in a number of disks \citep{oberg15b,huang16,oberg17,carney17,carney18}, but a test of the proposed mechanisms would require higher angular resolution observations of multiple molecules and detailed modeling.

\section{Disk chemistry and dust properties}


The H$_2$CO $3_{1,2} - 2_{1,1}$ ring is asymmetric as the redshifted side is $\sim 1.5$ times brighter than the blueshifted side (Figs.~\ref{fig:moments},  \ref{fig:mom8}, and \ref{fig:1dspec}).
The same asymmetry is observed in the map of the linearly polarized intensity of the continuum at 0.87~mm by \citet{bacciotti18}, where the redshifted disk side is brighter than the blueshifted side in the outer $40-50$ au region. The map by \citet{bacciotti18} also indicates that the orientation of the linear polarization vector changes from parallel to the disk minor axis in the inner disk region to azimuthal in the outer $40-50$ au. 
Interestingly, the radius where the change in the orientation of polarization occurs is coincident with the inner radius of the H$_2$CO ring.
The intensity and orientation of the polarization of the dust continuum caused by self-scattering strongly depends on the dust size, opacity, and degree of settling \citep{yang17}.
This suggests that the H$_2$CO chemistry is closely connected to the dust properties in the outer disk.
For example, a drop in the dust opacity beyond 40 au could be responsible for the change in polarization \citep{bacciotti18}.
If  the opacity of small grains also decreases, this would cause an enhancement of H$_2$CO desorbed from grains, due to increased UV and X-ray penetration. 
In the case of DG Tau, the desorption mechanism could be also favored by UV and X-rays produced in the blueshifted jet shocks \citep{coffey07,gudel08} that directly illuminate the disk from above.
The chemical and polarization change at 40 au could also be linked to the existence of substructures in the dust distribution.
The radial profile of the continuum intensity along the projected disk major axis shows a shoulder of emission at about 40~au (see Fig.~\ref{fig:cont}), which supports this hypothesis.

Finally, it is interesting to note that the H$_2$CO ring is offset by $\sim 0\farcs06$ ($\sim 7$ au) with respect to the dust continuum peak, i.e., the redshifted ring side is located farther from the continuum peak than the blueshifted side (Fig.~\ref{fig:moments}).
This also suggests an asymmetric distribution of the chemically enriched gas in the outer disk region.
However, higher angular resolution observations are needed to confirm this scenario. 

\section{Conclusions}

ALMA observations at $0\farcs15$ resolution of the disk of DG Tau show a ring of formaldehyde peaking at $\sim 62$ au, i.e., outside the CO iceline (R$_{\rm CO} \sim 30$~au) and close to the edge of the 1.3~mm dust continuum (R$_{\rm dust} \sim 66$ au). This suggests an enhancement of the H$_2$CO abundance in the outer disk due to H$_2$CO formation by CO hydrogenation on the icy grains. Moreover, the desorption of H$_2$CO from the grains could be enhanced at the edge of the mm dust continuum due to increased UV penetration and/or temperature inversion. 
The abundance ratio between CH$_3$OH and H$_2$CO is $< 1,$ in agreement with the predictions of disk chemistry models.
Interestingly, the inner edge of the H$_2$CO ring is at $\sim 40$ au, where the polarization of the dust continuum changes orientation and the continuum intensity shows a local enhancement, hinting at a tight link between the  H$_2$CO chemistry and the dust properties in the outer disk and at the possible presence of substructures in the dust distribution.

\begin{acknowledgements}
     This paper uses ALMA data from project 2016.1.00846.S. ALMA is a partnership of ESO (representing its member states), NSF (USA), and NINS (Japan), together with NRC (Canada), MOST and ASIAA (Taiwan), and KASI (Republic of Korea), in cooperation with the Republic of Chile. The Joint ALMA Observatory is operated by ESO, AUI/NRAO, and NAOJ. This work was partly supported by PRIN-INAF/2016 GENESIS-SKA and by the Italian Ministero dell'Istruzione, Universit\`a e Ricerca through the grants Progetti Premiali 2012/iALMA (CUP-C52I13000140001), 2017/FRONTIERA (CUP-C61I15000000001), and SIR-(RBSI14ZRHR). GG acknowledges the financial support of the Swiss National Science Foundation within the framework of the National Centre for Competence in Research PlanetS.
\end{acknowledgements}

   \bibliographystyle{aa} 
   \bibliography{../../mybibtex.bib} 

\begin{thebibliography}{49}
\expandafter\ifx\csname natexlab\endcsname\relax\def\natexlab#1{#1}\fi

\bibitem[{{Aikawa} \& {Herbst}(1999)}]{aikawa99}
{Aikawa}, Y. \& {Herbst}, E. 1999, \aap, 351, 233

\bibitem[{{Bacciotti} {et~al.}(2018){Bacciotti}, {Girart}, {Padovani}, {Podio},
  {Paladino}, {Testi}, {Bianchi}, {Galli}, {Codella}, {Coffey}, {Favre}, \&
  {Fedele}}]{bacciotti18}
{Bacciotti}, F., {Girart}, J.~M., {Padovani}, M., {et~al.} 2018, \apjl, 865,
  L12

\bibitem[{{Balucani} {et~al.}(2015){Balucani}, {Ceccarelli}, \&
  {Taquet}}]{balucani15}
{Balucani}, N., {Ceccarelli}, C., \& {Taquet}, V. 2015, \mnras, 449, L16

\bibitem[{{Bianchi} {et~al.}(2017){Bianchi}, {Codella}, {Ceccarelli},
  {Fontani}, {Testi}, {Bachiller}, {Lefloch}, {Podio}, \&
  {Taquet}}]{bianchi17a}
{Bianchi}, E., {Codella}, C., {Ceccarelli}, C., {et~al.} 2017, \mnras, 467,
  3011

\bibitem[{{Carney} {et~al.}(2018){Carney}, {Fedele}, {Hogerheijde}, {Favre},
  {Walsh}, {Bruderer}, {Miotello}, {Murillo}, {Klaassen}, {Henning}, \& {van
  Dishoeck}}]{carney18}
{Carney}, M.~T., {Fedele}, D., {Hogerheijde}, M.~R., {et~al.} 2018, \aap, 614,
  A106

\bibitem[{{Carney} {et~al.}(2019){Carney}, {Hogerheijde}, {Guzm{\'a}n},
  {Walsh}, {{\"O}berg}, {Fayolle}, {Cleeves}, {Carpenter}, \& {Qi}}]{carney19}
{Carney}, M.~T., {Hogerheijde}, M.~R., {Guzm{\'a}n}, V.~V., {et~al.} 2019,
  arXiv e-prints [\eprint[arXiv]{1901.02689}]

\bibitem[{{Carney} {et~al.}(2017){Carney}, {Hogerheijde}, {Loomis}, {Salinas},
  {{\"O}berg}, {Qi}, \& {Wilner}}]{carney17}
{Carney}, M.~T., {Hogerheijde}, M.~R., {Loomis}, R.~A., {et~al.} 2017, \aap,
  605, A21

\bibitem[{{Caselli} \& {Ceccarelli}(2012)}]{caselli12a}
{Caselli}, P. \& {Ceccarelli}, C. 2012, \aapr, 20, 56

\bibitem[{{Cleeves}(2016)}]{cleeves16}
{Cleeves}, L.~I. 2016, \apjl, 816, L21

\bibitem[{{Coffey} {et~al.}(2007){Coffey}, {Bacciotti}, {Ray}, {Eisl{\"o}ffel},
  \& {Woitas}}]{coffey07}
{Coffey}, D., {Bacciotti}, F., {Ray}, T.~P., {Eisl{\"o}ffel}, J., \& {Woitas},
  J. 2007, \apj, 663, 350

\bibitem[{{Eisl{\"o}ffel} \& {Mundt}(1998)}]{eisloffel98}
{Eisl{\"o}ffel}, J. \& {Mundt}, R. 1998, \aj, 115, 1554

\bibitem[{{Favre} {et~al.}(2018){Favre}, {Fedele}, {Semenov}, {Parfenov},
  {Codella}, {Ceccarelli}, {Bergin}, {Chapillon}, {Testi}, {Hersant},
  {Lefloch}, {Fontani}, {Blake}, {Cleeves}, {Qi}, {Schwarz}, \&
  {Taquet}}]{favre18}
{Favre}, C., {Fedele}, D., {Semenov}, D., {et~al.} 2018, \apjl, 862, L2

\bibitem[{{Fedele} {et~al.}(2013){Fedele}, {Bruderer}, {van Dishoeck}, {Carr},
  {Herczeg}, {Salyk}, {Evans}, {Bouwman}, {Meeus}, {Henning}, {Green},
  {Najita}, \& {G{\"u}del}}]{fedele13}
{Fedele}, D., {Bruderer}, S., {van Dishoeck}, E.~F., {et~al.} 2013, \aap, 559,
  A77

\bibitem[{{Gaia Collaboration} {et~al.}(2018){Gaia Collaboration}, {Brown},
  {Vallenari}, {Prusti}, {de Bruijne}, {Babusiaux}, {Bailer-Jones}, {Biermann},
  {Evans}, {Eyer}, \& et~al.}]{gaia18}
{Gaia Collaboration}, {Brown}, A.~G.~A., {Vallenari}, A., {et~al.} 2018, \aap,
  616, A1

\bibitem[{{Gaia Collaboration} {et~al.}(2016){Gaia Collaboration}, {Prusti},
  {de Bruijne}, {Brown}, {Vallenari}, {Babusiaux}, {Bailer-Jones}, {Bastian},
  {Biermann}, {Evans}, \& et~al.}]{gaia16}
{Gaia Collaboration}, {Prusti}, T., {de Bruijne}, J.~H.~J., {et~al.} 2016,
  \aap, 595, A1

\bibitem[{{Garrod} {et~al.}(2008){Garrod}, {Widicus Weaver}, \&
  {Herbst}}]{garrod08}
{Garrod}, R.~T., {Widicus Weaver}, S.~L., \& {Herbst}, E. 2008, \apj, 682, 283

\bibitem[{{G{\"u}del} {et~al.}(2018){G{\"u}del}, {Eibensteiner}, {Dionatos},
  {Audard}, {Forbrich}, {Kraus}, {Rab}, {Schneider}, {Skinner}, \&
  {Vorobyov}}]{gudel18}
{G{\"u}del}, M., {Eibensteiner}, C., {Dionatos}, O., {et~al.} 2018, \aap, 620,
  L1

\bibitem[{{G{\"u}del} {et~al.}(2008){G{\"u}del}, {Skinner}, {Audard}, {Briggs},
  \& {Cabrit}}]{gudel08}
{G{\"u}del}, M., {Skinner}, S.~L., {Audard}, M., {Briggs}, K.~R., \& {Cabrit},
  S. 2008, \aap, 478, 797

\bibitem[{{Guilloteau} {et~al.}(2013){Guilloteau}, {Di Folco}, {Dutrey},
  {Simon}, {Grosso}, \& {Pi{\'e}tu}}]{guilloteau13}
{Guilloteau}, S., {Di Folco}, E., {Dutrey}, A., {et~al.} 2013, \aap, 549, A92

\bibitem[{{Guzm{\'a}n} {et~al.}(2018){Guzm{\'a}n}, {{\"O}berg}, {Carpenter},
  {Le Gal}, {Qi}, \& {Pagues}}]{guzman18a}
{Guzm{\'a}n}, V.~V., {{\"O}berg}, K.~I., {Carpenter}, J., {et~al.} 2018, \apj,
  864, 170

\bibitem[{{Huang} {et~al.}(2016){Huang}, {{\"O}berg}, \& {Andrews}}]{huang16}
{Huang}, J., {{\"O}berg}, K.~I., \& {Andrews}, S.~M. 2016, \apjl, 823, L18

\bibitem[{{Isella} {et~al.}(2010){Isella}, {Natta}, {Wilner}, {Carpenter}, \&
  {Testi}}]{isella10}
{Isella}, A., {Natta}, A., {Wilner}, D., {Carpenter}, J.~M., \& {Testi}, L.
  2010, \apj, 725, 1735

\bibitem[{{Kamp} {et~al.}(2010){Kamp}, {Tilling}, {Woitke}, {Thi}, \&
  {Hogerheijde}}]{kamp10}
{Kamp}, I., {Tilling}, I., {Woitke}, P., {Thi}, W.-F., \& {Hogerheijde}, M.
  2010, \aap, 510, A18

\bibitem[{{Kitamura} {et~al.}(1996){Kitamura}, {Kawabe}, \&
  {Saito}}]{kitamura96a}
{Kitamura}, Y., {Kawabe}, R., \& {Saito}, M. 1996, \apj, 457, 277

\bibitem[{{Loomis} {et~al.}(2015){Loomis}, {Cleeves}, {{\"O}berg}, {Guzman}, \&
  {Andrews}}]{loomis15}
{Loomis}, R.~A., {Cleeves}, L.~I., {{\"O}berg}, K.~I., {Guzman}, V.~V., \&
  {Andrews}, S.~M. 2015, \apjl, 809, L25

\bibitem[{{Millar} {et~al.}(1991){Millar}, {Herbst}, \& {Charnley}}]{millar91}
{Millar}, T.~J., {Herbst}, E., \& {Charnley}, S.~B. 1991, \apj, 369, 147

\bibitem[{{M{\"u}ller} {et~al.}(2001){M{\"u}ller}, {Thorwirth}, {Roth}, \&
  {Winnewisser}}]{muller01}
{M{\"u}ller}, H.~S.~P., {Thorwirth}, S., {Roth}, D.~A., \& {Winnewisser}, G.
  2001, \aap, 370, L49

\bibitem[{{{\"O}berg} {et~al.}(2015{\natexlab{a}}){{\"O}berg}, {Furuya},
  {Loomis}, {Aikawa}, {Andrews}, {Qi}, {van Dishoeck}, \& {Wilner}}]{oberg15b}
{{\"O}berg}, K.~I., {Furuya}, K., {Loomis}, R., {et~al.} 2015{\natexlab{a}},
  \apj, 810, 112

\bibitem[{{{\"O}berg} {et~al.}(2015{\natexlab{b}}){{\"O}berg}, {Guzm{\'a}n},
  {Furuya}, {Qi}, {Aikawa}, {Andrews}, {Loomis}, \& {Wilner}}]{oberg15}
{{\"O}berg}, K.~I., {Guzm{\'a}n}, V.~V., {Furuya}, K., {et~al.}
  2015{\natexlab{b}}, \nat, 520, 198

\bibitem[{{{\"O}berg} {et~al.}(2017){{\"O}berg}, {Guzm{\'a}n}, {Merchantz},
  {Qi}, {Andrews}, {Cleeves}, {Huang}, {Loomis}, {Wilner}, {Brinch}, \&
  {Hogerheijde}}]{oberg17}
{{\"O}berg}, K.~I., {Guzm{\'a}n}, V.~V., {Merchantz}, C.~J., {et~al.} 2017,
  \apj, 839, 43

\bibitem[{{{\"O}berg} {et~al.}(2010){{\"O}berg}, {Qi}, {Fogel}, {Bergin},
  {Andrews}, {Espaillat}, {van Kempen}, {Wilner}, \& {Pascucci}}]{oberg10}
{{\"O}berg}, K.~I., {Qi}, C., {Fogel}, J.~K.~J., {et~al.} 2010, \apj, 720, 480

\bibitem[{{{\"O}berg} {et~al.}(2011){{\"O}berg}, {Qi}, {Fogel}, {Bergin},
  {Andrews}, {Espaillat}, {Wilner}, {Pascucci}, \& {Kastner}}]{oberg11}
{{\"O}berg}, K.~I., {Qi}, C., {Fogel}, J.~K.~J., {et~al.} 2011, \apj, 734, 98

\bibitem[{{Pacheco-V{\'a}zquez} {et~al.}(2016){Pacheco-V{\'a}zquez}, {Fuente},
  {Baruteau}, {Bern{\'e}}, {Ag{\'u}ndez}, {Neri}, {Goicoechea}, {Cernicharo},
  \& {Bachiller}}]{pachecovazquez16}
{Pacheco-V{\'a}zquez}, S., {Fuente}, A., {Baruteau}, C., {et~al.} 2016, \aap,
  589, A60

\bibitem[{{Podio} {et~al.}(2013){Podio}, {Kamp}, {Codella}, {Cabrit}, {Nisini},
  {Dougados}, {Sandell}, {Williams}, {Testi}, {Thi}, {Woitke}, {Meijerink},
  {Spaans}, {Aresu}, {M{\'e}nard}, \& {Pinte}}]{podio13}
{Podio}, L., {Kamp}, I., {Codella}, C., {et~al.} 2013, \apjl, 766, L5

\bibitem[{{Podio} {et~al.}(2012){Podio}, {Kamp}, {Flower}, {Howard}, {Sandell},
  {Mora}, {Aresu}, {Brittain}, {Dent}, {Pinte}, \& {White}}]{podio12}
{Podio}, L., {Kamp}, I., {Flower}, D., {et~al.} 2012, \aap, 545, A44

\bibitem[{{Qi} {et~al.}(2013){Qi}, {{\"O}berg}, \& {Wilner}}]{qi13a}
{Qi}, C., {{\"O}berg}, K.~I., \& {Wilner}, D.~J. 2013, \apj, 765, 34

\bibitem[{{Schuster} {et~al.}(1993){Schuster}, {Harris}, {Anderson}, \&
  {Russell}}]{schuster93}
{Schuster}, K.~F., {Harris}, A.~I., {Anderson}, N., \& {Russell}, A.~P.~G.
  1993, \apjl, 412, L67

\bibitem[{{Shirley}(2015)}]{shirley15}
{Shirley}, Y.~L. 2015, \pasp, 127, 299

\bibitem[{{Siess} {et~al.}(2000){Siess}, {Dufour}, \& {Forestini}}]{siess00}
{Siess}, L., {Dufour}, E., \& {Forestini}, M. 2000, \aap, 358, 593

\bibitem[{{Testi} {et~al.}(2002){Testi}, {Bacciotti}, {Sargent}, {Ray}, \&
  {Eisl{\"o}ffel}}]{testi02}
{Testi}, L., {Bacciotti}, F., {Sargent}, A.~I., {Ray}, T.~P., \&
  {Eisl{\"o}ffel}, J. 2002, \aap, 394, L31

\bibitem[{{Tielens} \& {Hagen}(1982)}]{tielens82}
{Tielens}, A.~G.~G.~M. \& {Hagen}, W. 1982, \aap, 114, 245

\bibitem[{{van der Marel} {et~al.}(2014){van der Marel}, {van Dishoeck},
  {Bruderer}, \& {van Kempen}}]{vandermarel14}
{van der Marel}, N., {van Dishoeck}, E.~F., {Bruderer}, S., \& {van Kempen},
  T.~A. 2014, \aap, 563, A113

\bibitem[{{van der Tak} {et~al.}(2007){van der Tak}, {Black}, {Sch{\"o}ier},
  {Jansen}, \& {van Dishoeck}}]{vandertak07}
{van der Tak}, F.~F.~S., {Black}, J.~H., {Sch{\"o}ier}, F.~L., {Jansen}, D.~J.,
  \& {van Dishoeck}, E.~F. 2007, \aap, 468, 627

\bibitem[{{van 't Hoff} {et~al.}(2018){van 't Hoff}, {Tobin}, {Trapman},
  {Harsono}, {Sheehan}, {Fischer}, {Megeath}, \& {van Dishoeck}}]{vanthoff18}
{van 't Hoff}, M.~L.~R., {Tobin}, J.~J., {Trapman}, L., {et~al.} 2018, \apjl,
  864, L23

\bibitem[{{Walsh} {et~al.}(2016){Walsh}, {Loomis}, {{\"O}berg}, {Kama}, {van 't
  Hoff}, {Millar}, {Aikawa}, {Herbst}, {Widicus Weaver}, \& {Nomura}}]{walsh16}
{Walsh}, C., {Loomis}, R.~A., {{\"O}berg}, K.~I., {et~al.} 2016, \apjl, 823,
  L10

\bibitem[{{Walsh} {et~al.}(2014){Walsh}, {Millar}, {Nomura}, {Herbst}, {Widicus
  Weaver}, {Aikawa}, {Laas}, \& {Vasyunin}}]{walsh14}
{Walsh}, C., {Millar}, T.~J., {Nomura}, H., {et~al.} 2014, \aap, 563, A33

\bibitem[{{Watanabe} \& {Kouchi}(2002)}]{watanabe02}
{Watanabe}, N. \& {Kouchi}, A. 2002, \apjl, 571, L173

\bibitem[{{Willacy} \& {Woods}(2009)}]{willacy09}
{Willacy}, K. \& {Woods}, P.~M. 2009, \apj, 703, 479

\bibitem[{{Yang} {et~al.}(2017){Yang}, {Li}, {Looney}, {Girart}, \&
  {Stephens}}]{yang17}
{Yang}, H., {Li}, Z.-Y., {Looney}, L.~W., {Girart}, J.~M., \& {Stephens}, I.~W.
  2017, \mnras, 472, 373

\end{thebibliography}

\begin{appendix}

\section{H$_2$CO channel maps, moment 8 map, and 1D spectrum}
\label{app:channel_maps}

\begin{figure*}
   \centering
   \includegraphics[width=17.cm]{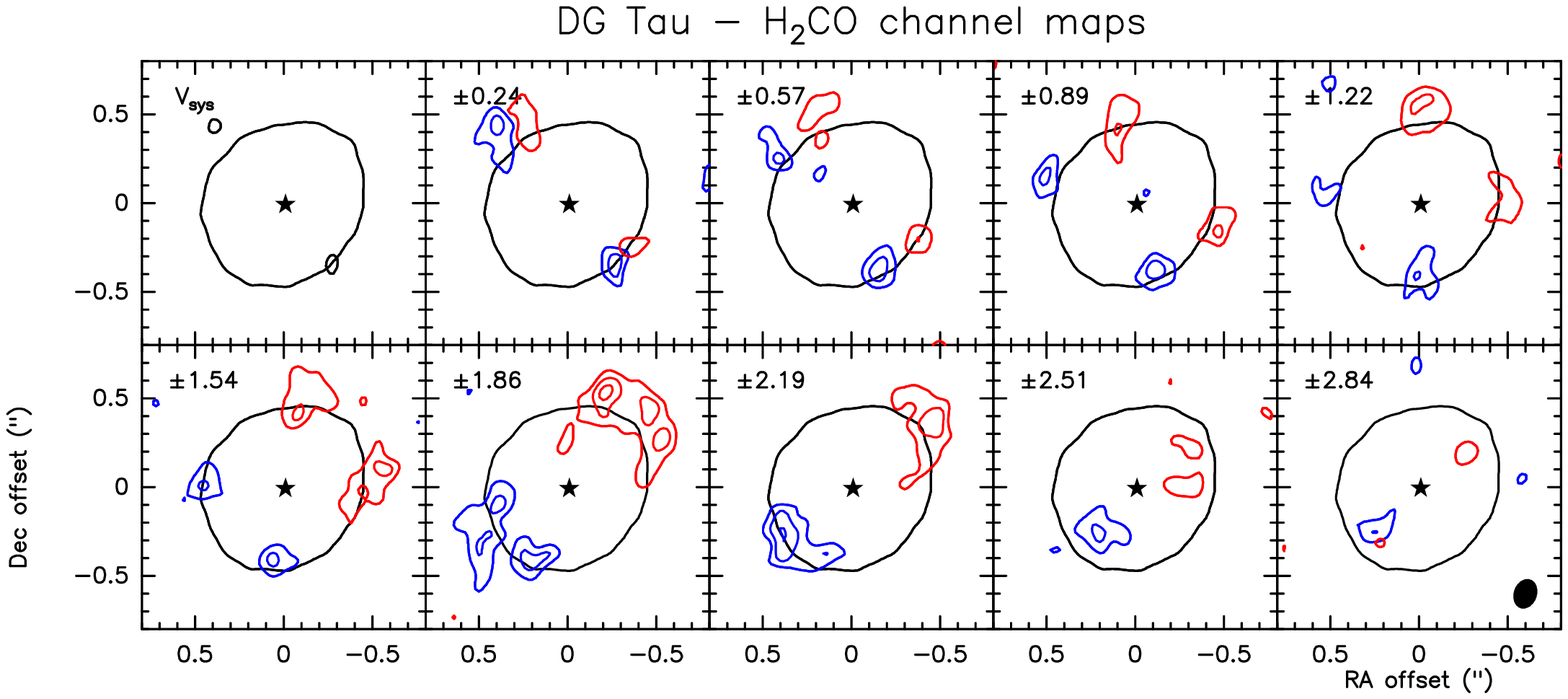}
   \caption{Channel maps of H$_2$CO $3_{12}-2_{11}$ towards DG Tau. The blue and red contours show the emission at symmetric blue- and redshifted velocities with respect to systemic (V$_{\rm sys}=+6.24$ \kms), as labeled in the upper right corner (V-V$_{\rm sys}$ in \kms). The first contour is at 5$\sigma$ with steps of 3$\sigma$. The black star and contour indicates the peak and the 5$\sigma$ level of the 1.3~mm continuum. The ellipse in the bottom right corner of the last channel map shows the ALMA synthesized beam.}
              \label{fig:channel_maps}%
    \end{figure*}

\begin{figure}
   \centering
   \includegraphics[width=8.5cm]{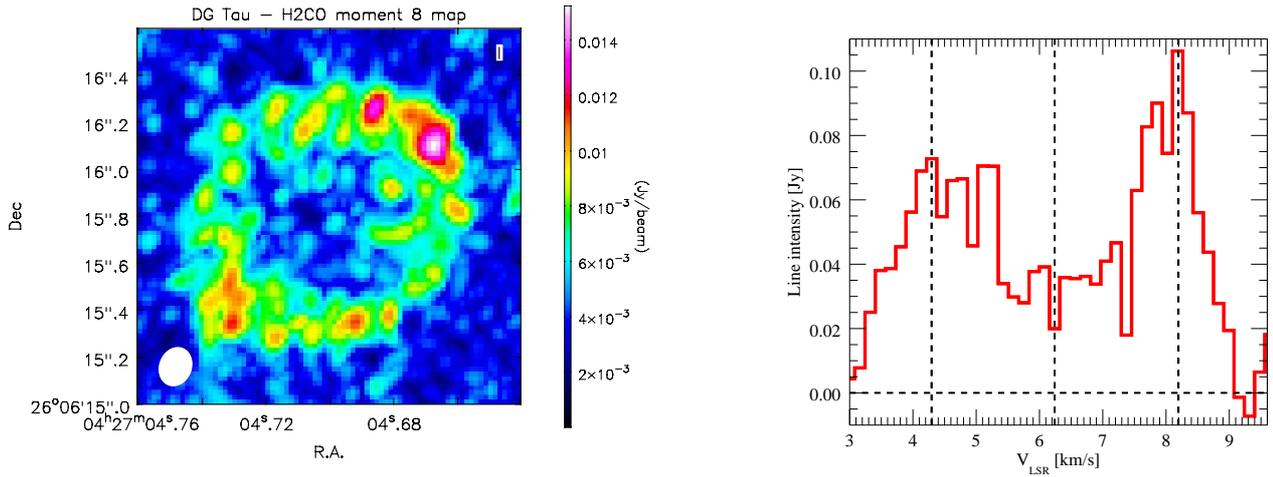} 
   \caption{H$_2$CO moment 8 map towards DG Tau. The color scale indicates the line intensity in Jy/beam. The white ellipse in the bottom left corner indicates the synthesized beam.} 
              \label{fig:mom8}
    \end{figure}

The channel maps and the moment 8 map obtained from the H$_2$CO $3_{12}-2_{11}$ line cube towards DG Tau are shown in Fig.~\ref{fig:channel_maps} and \ref{fig:mom8}.
The moment 8 map strongly supports the ring detection with a S/N $\sim 5$.

\begin{figure}
   \centering
   \includegraphics[width=7.cm]{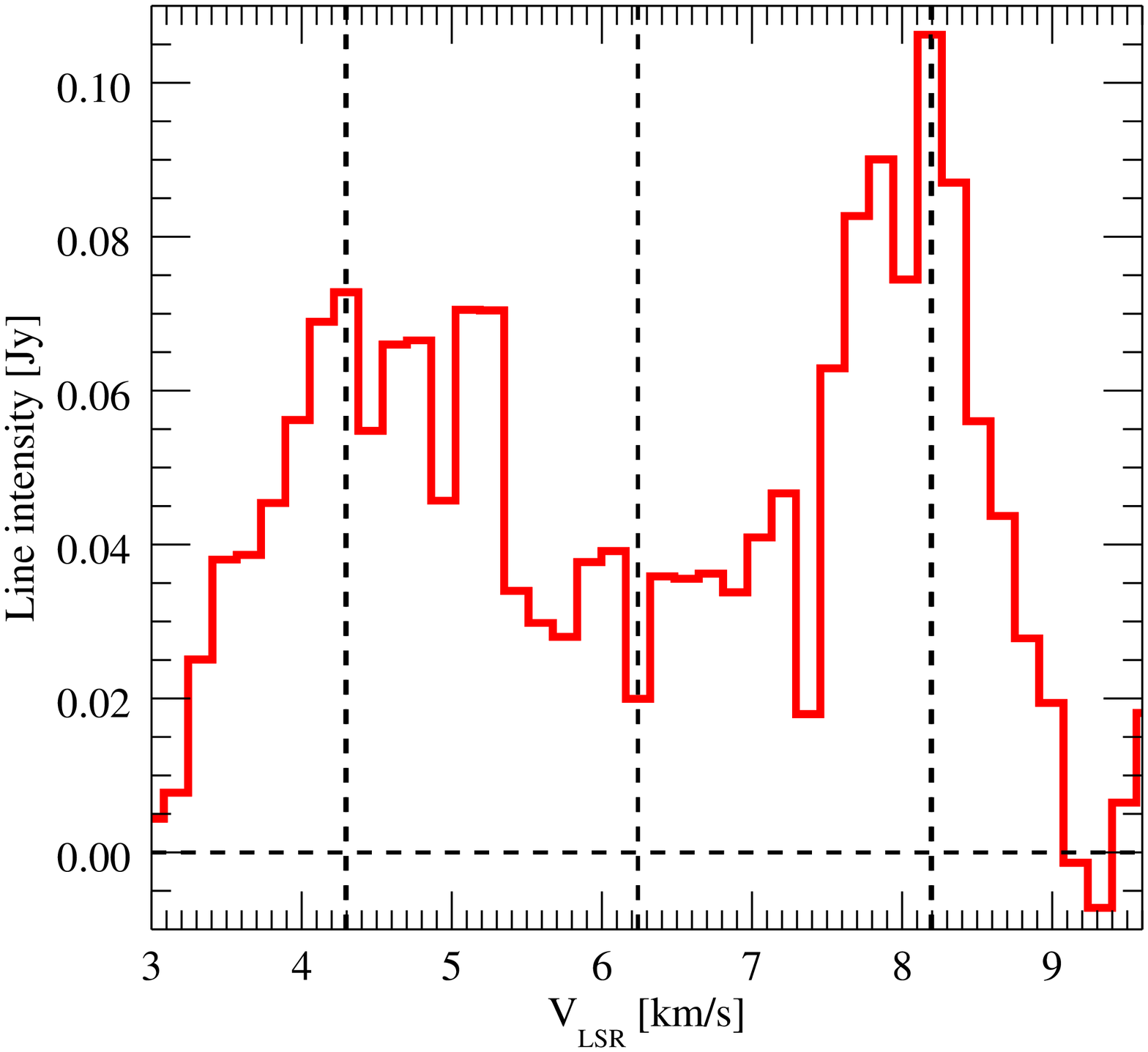}
   \caption{H$_2$CO $3_{1,2}-2_{1,1}$ spectrum integrated over a $0\farcs33-0\farcs75$ ring area. The vertical dashed lines indicate the systemic velocity, V$_{\rm sys} = +6.24$ \kms, and the position of the blue- and redshifted peaks ($V_{\rm peak} = V_{\rm sys} \pm 1.95$ \kms).} 
              \label{fig:1dspec}
    \end{figure}

The 1D spectrum obtained integrating the H$_2$CO line cube over a circular ring extending from $0\farcs33$ to $0\farcs75$ is shown in Fig.~\ref{fig:1dspec}. 

\section{Analysis of the continuum radial intensity profile}
\label{app:cont}

\begin{figure}
   \centering
   \hspace{-1.cm}
   \includegraphics[width=6.cm, angle=90]{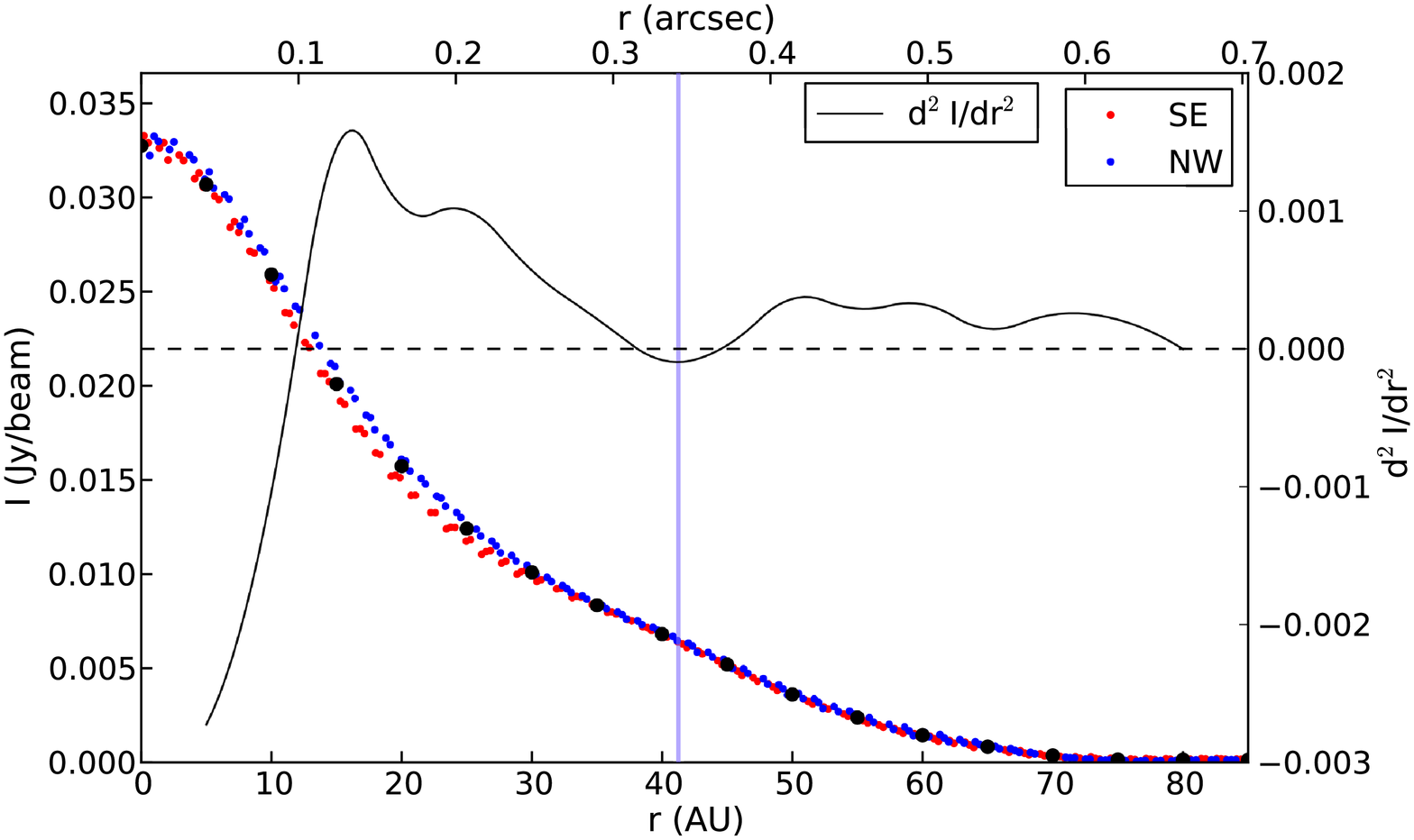}
   \caption{Radial cut of the continuum intensity map of DG~Tau at 1.3~mm: every point corresponds to a pixel inside a half-beam across the disk major axis ($i= 41\degr$, PA $= 128\degr$). Red dots represent the southeast side of the disk, blue dots the northwest side. The large black  dots correspond to 5~au bins computed by averaging  the two sides. The black curve is the second derivative of the continuum intensity profile. The solid vertical line indicates the center of the continuum enhancement, at 41 au.}
              \label{fig:cont}%
    \end{figure}

In Fig. \ref{fig:cont} we plot the radial profile of the continuum intensity along the projected disk major axis, in order to achieve the highest resolution, and avoid the dilution of possible substructures in the dust flux that could result from azimuthal averaging. 
No strong asymmetries are visible between the northwest and the southeast sides of the disk. A tentative shoulder appears at about 40~au from the star: to characterize its position we computed the second derivative of the radial intensity profile (by calculating the differentials over bins of 5~au, averaged between the the NW and SE sides). The radii at which the second derivative becomes negative give us the locations where there is a local increase in the continuum intensity and the profile becomes locally convex. This occurs in the central region ($<11.5$~au), corresponding to the peak of the continuum, and in the range between 38 and 44.5~au where the small shoulder is visible in the intensity profile. Hence, the center of the observed continuum intensity enhancement is located at $\sim41$~au.

\end{appendix}

\end{document}